\documentclass[review,12pt,3p]{elsarticle}

\usepackage{lineno,hyperref}
\modulolinenumbers[5]

\journal{Carbon}

\usepackage{xcolor}

\begin{document}

\title{Suppression of coherent thermal transport in quasiperiodic graphene-hBN superlattice ribbons}

\author[ufrn]{Isaac M. Felix}
\author[ufrn,ufpe]{Luiz Felipe C. Pereira\corref{cor1}}
\ead{pereira@df.ufpe.br}
\address[ufrn]{Departamento de F\'{\i}sica, Universidade Federal do Rio Grande do Norte, Natal, 59078-970, Brazil}
\address[ufpe]{Departamento de F\'{\i}sica, Universidade Federal de Pernambuco, Recife, 50670-901, Brazil}

\cortext[cor1]{Corresponding author}

\date{\today}

\begin{abstract}
Nanostructured superlattices are promising materials for novel electronic devices due to their adjustable physical properties. Periodic superlattices facilitate coherent phonon thermal transport due to constructive wave interference at the boundaries between the materials. However, it is possible to induce a crossover from coherent to incoherent transport regimes by adjusting the superlattice period. We have recently observed such crossover in periodic graphene-boron nitride nanoribbons as the length of individual domains was increased. In general, transport properties are dominated by translational symmetry and the presence of unconventional symmetries leads to unusual transport characteristics. Here we perform non-equilibrium molecular dynamics simulations to investigate phonon heat transport in graphene-hBN superlattices following the Fibonacci quasiperiodic sequence, which lie between periodic and disordered structures. Our simulations show that the quasiperiodicity can suppress coherent phonon thermal transport in these superlattices. This behavior is related to the increasing number of interfaces per unit cell as the Fibonacci generation increases, hindering phonon coherence along the superlattice. The suppression of coherent thermal transport in graphene-hBN superlattices enables a higher degree of control on heat conduction at the nanoscale, and shows potential for application in the design of novel thermal management devices.
\end{abstract}


\maketitle

\section{Introduction}

The term superlattice usually refers to a periodic structure of layers composed of two or more materials. 
The physical properties of a superlattice can be quite different from the ones of its constituents, and also originate unusual phenomena.
For example, in the realm of heat transport, periodic superlattices enable coherent propagation of phonons due to constructive wave interference at the interfaces between the materials.
Indeed, recent experiments in GaAs/AlAs superlattices \cite{Luckyanova2012,Luckyanova2018}, perovskite oxide superlattices \cite{Ravichandran2014} and TiN/(Al,Sc)N superlattices \cite{saha2017}, have shown that coherent phonons dominate their heat transport characteristics \cite{simkin2000, maldovan2015}. 
{Molecular dynamics simulations on superlattices of graphene-hBN \cite{zhu2014, da2016, chen2016, felix2018}, Si-Ge \cite{Chen2009,hu2012,lin2013,mu2015.2,kandemir2017},  silicene-germanene \cite{wang2017b}, graphene-nitrogenated holey graphene \cite{wang2017a}, isotopically modified graphene \cite{mu2015}, among others \cite{chen2005, Latour2014, frieling2014, bracht2014, xiong2015, mizuno2015, chakraborty2017, termentzidis2018} also corroborate the presence of coherent phonon transport in periodic superlattices and phononic crystals \cite{Hu2018,Hu2019}. }
In those studies, the thermal conductivity of the superlattice is always lower than the one of the individual pristine materials. 
{
Increasing the scattering of coherent phonons or preventing their formation is an efficient mechanism to further decrease the lattice thermal conductivity of superlattices, and enable their application in thermal rectification and thermoelectric devices \cite{Liu2014a, Chen2018, Zhang2018c, Chen2019a, Zeng2019}.
}
Nonetheless, deviations from periodicity can originate even more complex phenomena and unusual physical effects. 

The discovery of quasicrystals spawned a new area in solid state physics and chemistry \cite{Shechtman1984, levine1984}.
One of the first quasiperiodic superlattices consisted of a GaAs/AlAs structure following the Fibonacci sequence \cite{merlin1985} .
Quasiperiodic superlattices are generally formed by two (or more) types of domains, and can be seen as an intermediate between a periodic and a random arrangement \cite{albuquerque2004}.
Therefore, quasiperiodic patterning offers a possibility for tuning the physical properties of a given system in a controlled manner \cite{yu2006,zhao2011,lang2012}.
In fact, quasiperiodicity could lower the thermal conductivity of superlattices even further because phonons of various  wavelengths will be scattered by structures with different length scales. 
A similar behavior was recently observed in hierarchical superlattices \cite{mu2015.2, mu2015}.

Graphene-hBN monolayers have recently been synthesized via lithography patterning coupled with chemical vapor deposition \cite{ci2010, Liu2011, gao2013, Liu2013}. 
This approach enables the fabrication of large-scale hybrid graphene-hBN heterostructures which are continuous and  can be transferred to substrates \cite{Liu2013}. 
Furthermore, these materials have been shown to possess unusual physical properties, quite different from its constituents \cite{ding2009, ci2010, seol2011, shinde2011, Kan2011, Bernardi2012, jain2013}. 
In a recent work we employed molecular dynamics simulations to show that graphene-hBN superlattice ribbons can present a 98\% reduction in thermal conductivity when compared to pristine graphene, depending on the superlattice period \cite{felix2018, balandin2008, xu2014}. 
In addition, our results identified a transition from coherent (wave-like) to incoherent (particle-like) heat transport as the superlattice period was increased from a minimum value \cite{felix2018}.

In the present work we investigate the thermal conductivity of graphene-hBN superlattice ribbons via non-equilibrium molecular dynamics simulations, alternating equal domains of graphene and hBN according to the Fibonacci sequence.
Starting from the periodic superlattice we observe that the thermal conductivity reaches a minimum value as the length of the individual domains increase. 
We also notice that the domain length in which the thermal conductivity reaches a minimum  disappears as the Fibonacci generation increases.
Since the presence of a valley in the thermal conductivity is associated with a crossover from a coherent to an incoherent transport regime, we conclude that the presence of a quasiperiodicity supresses coherent phonon transport in graphene-hBN superlattices.
We attribute this behavior to the increasing number of interfaces per unit cell as the Fibonacci generation increases, which hinders phonon coherence along the superlattice.
This suppression of coherent thermal transport in graphene-hBN superlattices enables a further degree of control on heat conduction at the nanoscale, which has the potential to improve the design of novel thermal management devices.

\begin{figure*}[htbp]
\begin{center}
\includegraphics[width=\linewidth]{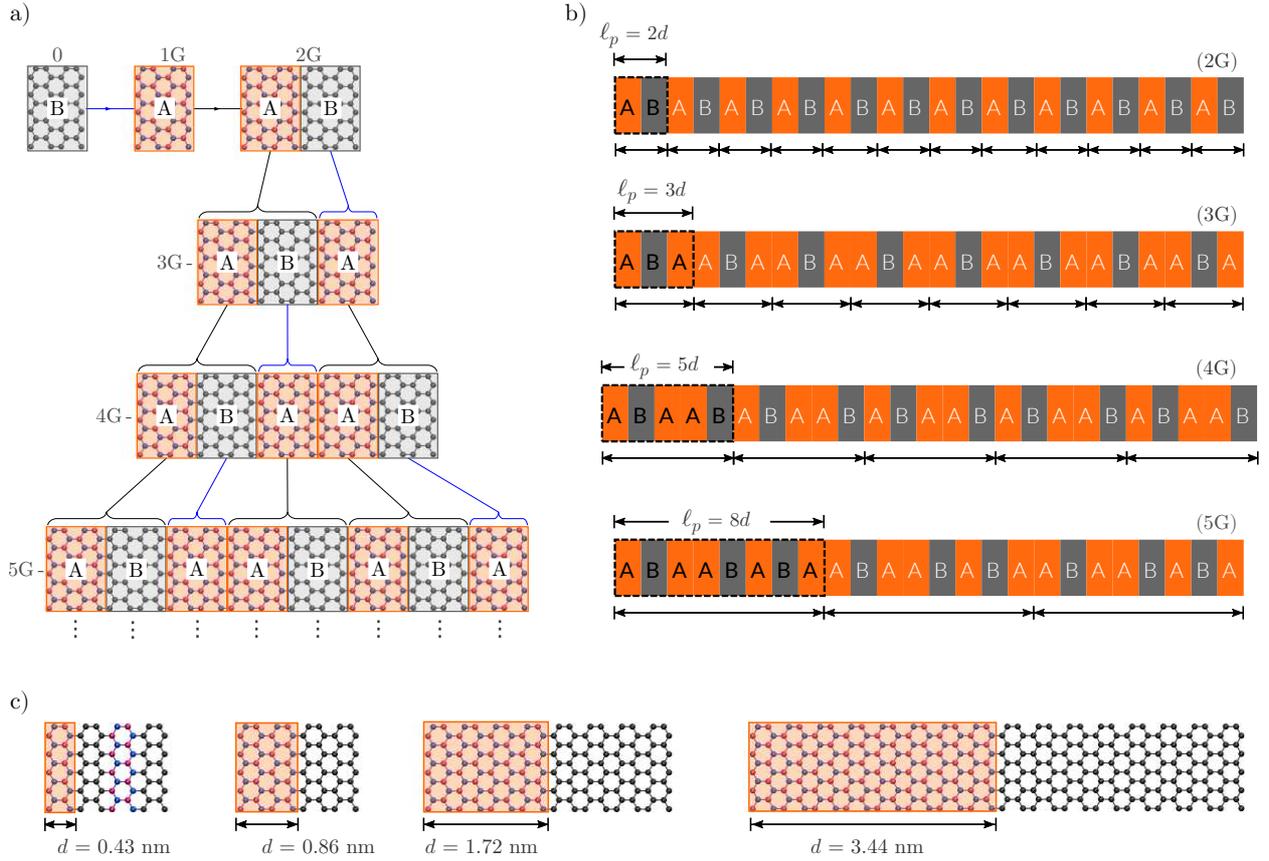}
\caption{(a) Unit cell of superlattices following the Fibonacci sequence up to fifth generation (5G). Domains of type $B$ correspond to graphene while type $A$ correspond to hBN. (b) Superlattice nanoribbons are obtained by replicating the supercell of each generation. (c) We consider superlattices with equal-sized domains of graphene and hBN, but vary the domain size $d$. All ribbons have a nominal width of $5$ nm.}
\label{fig01}
\end{center}
\end{figure*}

\section{Methods}

The Fibonacci sequence is an infinite sequence of integer numbers where each element is the sum the two preceding ones. 
Starting from 0 and 1, the sequence of Fibonacci numbers is: $1, 1, 2, 3, 5, 8, \dots$
In our case, we consider a sequence formed by two types of domains $A$ (hBN) and $B$ (graphene), such that each generation is obtained iteratively from its predecessors by an inflation rule $\mathrm{A}\rightarrow \mathrm{AB}$ and $\mathrm{B}\rightarrow \mathrm{A}$.
Therefore, the sequence reads: $B, A, AB, ABA, ABAAB, ABAABABA, \dots$, as illustrated in Fig. \ref{fig01}(a) up to the fifth generation (5G).
The number of individual domains in each generation grows rapidly and coincides with the respective Fibonacci number.
As the generation increases, the ratio between the number of domains of type $A$ and domains of type $B$ approaches the so-called golden ratio: $\frac{N_A}{N_B} \rightarrow (1+\sqrt{5})/2 \approx 1.618$.
Therefore, the number of domains of type $A$ is always lager than the number of domains of type $B$. 
Furthermore, domains of type $B$ are never consecutive, whereas two domains of type $A$ can be adjacent to each other.

We build graphene-hBN superlattice nanoribbons by repeating the unit cell of the respective Fibonacci generation, as illustrated in Fig. \ref{fig01}(b).
The second fibonacci generation (2G) is the periodic case which we investigated recently \cite{felix2018}.
The length of the supercell defines a superlattice period for each generation, which is given by $\ell_p = F_g \cdot d$, where $F_g$ is the Fibonacci number and $d$ is the length of the individual domains.
We have considered individual domains of various lengths but constrained the length of graphene domains to equal the length of hBN domains, as shown in Fig. \ref{fig01}(c).
Therefore, our samples have equal-sized domains of graphene and hBN for each Fibonacci generation.
Finally, all superlattice ribbons considered in this work have a fixed nominal width of $5$ nm. 

Molecular dynamics simulations were performed with LAMMPS (Large-scale Atomic/Molecular Massively Parallel Simulator) \cite{Plimpton1995}. 
{The interatomic forces were described by the Tersoff empirical potential optimized for carbon and BN nanostructures \cite{Tersoff1988,Lindsay2010,Sevik2011a,kinaci2012}, which has been thoroughly tested and employed in several C-BN systems.}
In all simulations the equations of motion were integrated with a $0.5$ fs timestep. 
The systems were initially thermalized with a Nos\'e-Hoover thermostat at $300$ K for $100$ ps. 
{We employed periodic boundary conditions along the length of the superlattice ribbons and free boundary conditions along the other two directions.}
Each ribbon was relaxed at finite temperature in order to achieve zero-stress along the periodic direction, the stress along the other two directions is automatically zero.
The thermostat was turned off after equilibration and the system was then allowed to evolve under microcanonical conditions.

\begin{figure}[htbp]
\begin{center}
\hspace{1.5cm}(a)\includegraphics[width=0.65\linewidth]{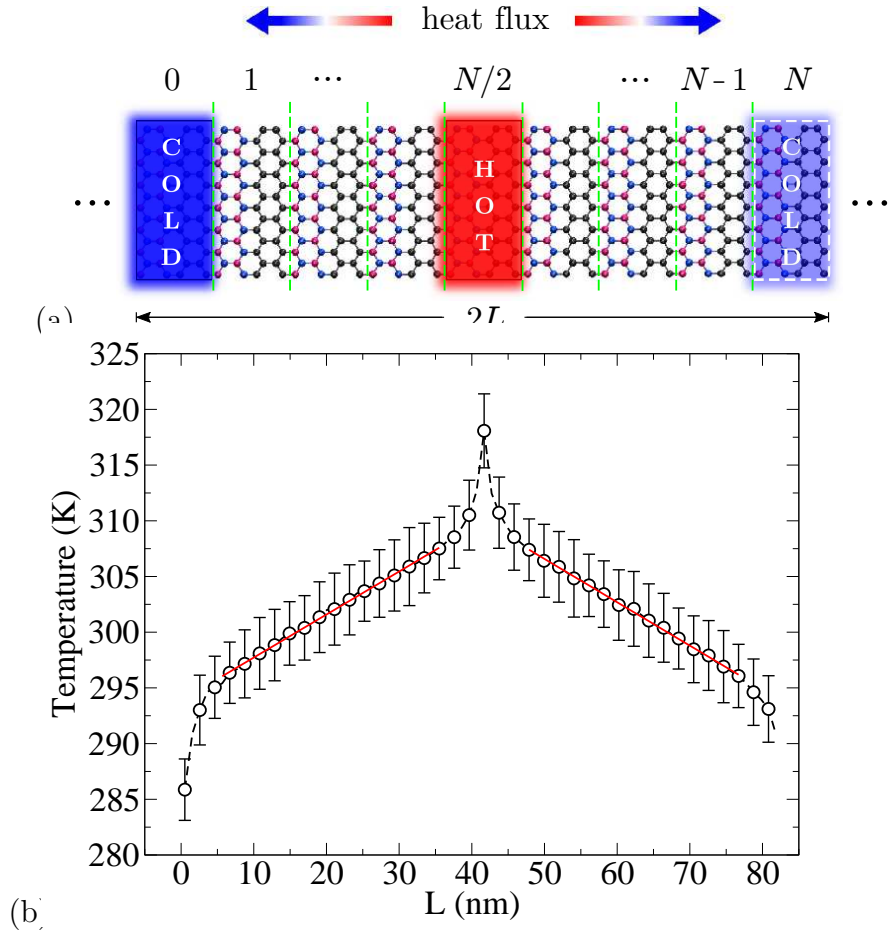}
(b)\includegraphics[width=0.6\linewidth]{fig02b.eps}
\caption{(a) In the RNEMD method the system is divided in $N$ slabs of equal length. Slab $0$ is the heat sink and slab $N/2$ is the heat source. Due to periodic boundary conditions, slab $N$ is the image of slab $0$.
(b) Typical temperature profile for a 5G superlattice with $d=0.43$ nm in the steady state.}
\label{fig02}
\end{center}
\end{figure}

The lattice (phonon) thermal conductivity was computed with the reverse non-equilibrium molecular dynamics (RNEMD) method \cite{Muller-Plathe1997}, which imposes a heat flux through the system by swapping the kinetic energy of atoms in two separate regions, originating a temperature gradient.
In the RNEMD method, the system is divided into $N$ slabs along its length and two of those slabs are chosen as heat source and heat sink.
Fast moving atoms in the heat sink have their kinetic energy swapped with slow moving atoms in the heat source, generating a temperature gradient.
In our setup we choose the first slab as the heat sink and the middle slab as the heat source, as shown in Fig. \ref{fig02}.
Each slab had an average of $200$ atoms.

The heat flux $J$ was calculated from the diference in kinetic energy of the exchanged particles 
\begin{equation}
J(t) =\frac{1}{2 \Delta t A} \; {\sum_{\mathrm{swaps}} \left[ K_{\mathrm{source}} - K_{\mathrm{sink}} \right] },
\end{equation}
where $\Delta t$ is the time interval measured from the beginning of the swaps, $K_\mathrm{source}$ and $K_\mathrm{sink}$ are the kinetic energies of the atoms in the heat source and sink, and the sum is over all swaps. 
We define the cross sectional area $A$ of the nanoribbon as the nominal width of $5$ nm multiplied by a thickness of $0.33$ nm.
The factor of $2$ accounts for the fact that the heat flux is divided between both sides of the heat source.
When the swaps are introduced the system is taken out of equilibrium, and as they proceed it eventually reaches a steady state where the time average of the heat flux is constant.
The frequency with which kinetic energy swaps take place has to be carefully adjusted. Every swap introduces a little discontinuity in the trajectory of the atoms involved. If the swaps happen too often the total energy of the system tends to increase. However, if steps happen too seldom the system takes too long to reach a steady state.
In our simulations we performed the energy swaps every $1000$ timesteps for a total simulation time of $40 \times 10^7$ timesteps, which corresponds to $20$ ns.

At every simulation step the temperature of each slab can be calculated from the equipartition theorem as
\begin{equation}
T_i = \frac{1}{3 n_i k_B}\sum\limits_{j=1}^{n_i}\frac{p_j^2}{m_j},
\end{equation}
where $T_i$ is the temperature of $i$-th slab, $n_i$ is the number of atoms in it, $k_B$ is Boltzmann's constant, $m_j$ is the mass of atom $j$, and $p_j$ stands for its linear momentum.
{
From the temperature of each slab we build a temperature profile for the system, as the one shown in Fig. \ref{fig02}(b), from which the temperature gradient can be measured as indicated by the red lines in the figure. 
}
Since we are dealing with superlattice nanoribbons, our systems are effectively one-dimensional and only one component of the temperature gradient is relevant.
Due to our geometry we have two estimates for the gradient, one from each side of the heat source, as illustrated in Fig. \ref{fig02}(b).
In fact, one way to estimate if the system has reached the steady state is to check if the gradients on each side of the heat source are equal (within error bars).

Once the system reaches the steady state, both the heat flux and the temperature gradient are stationary and we can calculate the lattice thermal conductivity for a system of length $L$ from Fourier law 
\begin{equation}
\kappa(L) =  \frac{\langle J \rangle}{ \left \langle {\partial T}/{\partial x}\right\rangle},
\label{eq:fourier}
\end{equation}
where $\langle \cdots \rangle$ represents the time average of the quantities, and $x$ is the direction of the heat flow.
In general, the behavior of $\kappa(L)$ can be described by a ballistic-to-diffusive expression such as \cite{schelling2002}
\begin{equation}
\label{eq:k-vs-L}
\frac{1}{\kappa(L)} = \frac{1}{\kappa} \left[ 1 + \frac{\Lambda}{L} \right]
\end{equation}
where $\kappa$ is the intrinsic (length independent) thermal conductivity of the material, and $\Lambda$ is an effective phonon mean free path (MFP) \cite{Pereira2016}.
Fan et al. have shown that for pristine graphene sheets the expression above is more accurate when $\kappa$ and $\Lambda$ are decomposed between in-plane and out-of-plane components \cite{Fan2017}.
However, we have verified that for graphene-hBN superlattice ribbons the decomposition is not necessary and Eq. (\ref{eq:k-vs-L}) provides accurate results \cite{felix2018}.
In what follows we employ Eq. (\ref{eq:k-vs-L}) to determine the influence of domain length and Fibonacci generation on the intrinsic thermal conductivity of graphene-hBN nanoribbons.

\section{Results and Discussion}

\begin{figure}[htbp]
\begin{center}
\includegraphics[width=0.8\linewidth]{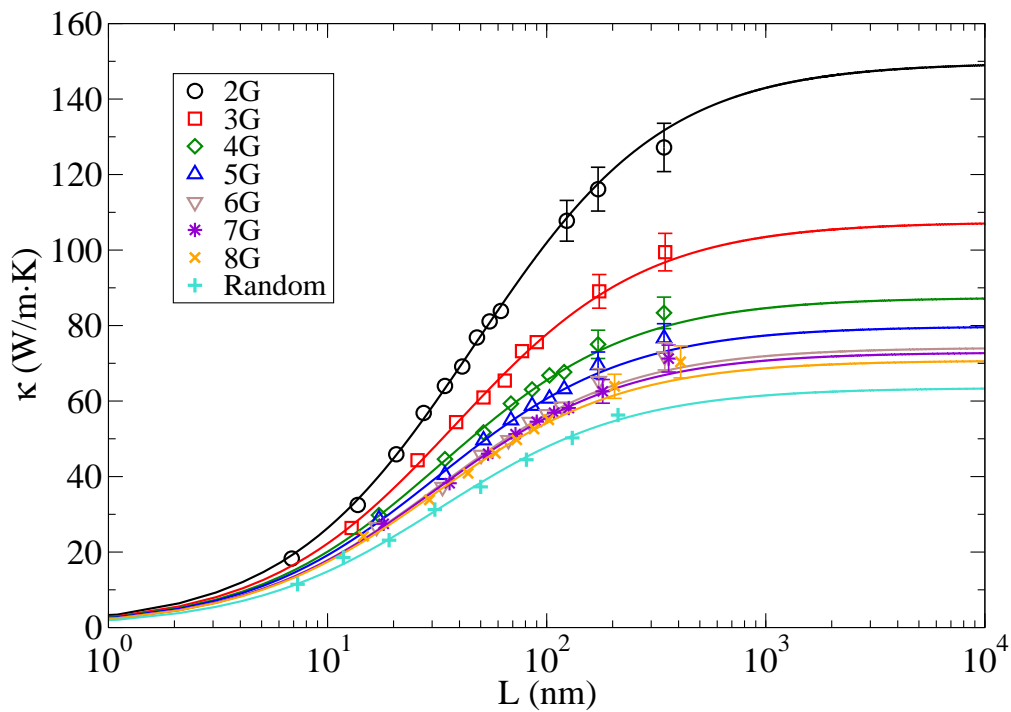}
\caption{Thermal conductivity versus nanoribbon length $L$ for domains of length $d = 0.43$ nm. Data points are extracted from RNEMD simulations and lines are least-square fits to the data. Notice that $\kappa (L)$ decreases with the Fibonacci generation and approaches the random superlattice.}
\label{fig03}
\end{center}
\end{figure}

{
We begin by analyzing the dependence of the thermal conductivity with the total length of the superlattice ribbons. 
At first we consider superlattices with a fixed domain length of $d = 0.43$ nm, which corresponds to the smallest domain size in Fig. \ref{fig01}(c), including only one hexagon of atoms per domain.
In Fig. \ref{fig03} we present the thermal conductivity as a function $L$ from the second to the eighth Fibonacci generation.
The figure also includes data for a random superlattice with the same domain length.
The random lattice was generated by randomly distributing domains of graphene and hBN subject to two constraints: domains of type $B$ (graphene) are never adjacent, and the ratio between $A$ and $B$ domains ($N_A/N_B$) is as close as possible to the golden ratio.
These constraints are necessary in order to compare the quasiperiodic superlattices to equivalent random lattices, since the thermal conductivity of graphene is much larger than hBN, and the superlattice conductivity is strongly dependent on the number of interfaces.
}

{
The symbols in Fig. \ref{fig03} represent values calculated from Eq. (\ref{eq:fourier}), while the continuous lines are least-squares fits to the data points using Eq. (\ref{eq:k-vs-L}).
In order to estimate the uncertainty in the calculated data points we performed independent simulations by changing the initial atomic velocities, and found an uncertainty smaller than 5\% in all cases.
We have also estimated the uncertainty in $\kappa$ within an individual simulation by considering the uncertainties in the average heat flux and temperature gradient. 
In this case we also found an uncertainty smaller than 5\%.
For the sake of clarity we omit the error bars for quasiperiodic superlattices with $L < 125$ nm and for the random lattice.
Furthermore, in order to assert the predictive power of the ballistic-to-diffusive expression, the fitting procedure did not include the three longest systems for the second generation and the two longest ones for generations 3 to 8.
Therefore, error bars representing a $5$\% uncertainty are shown only for the largest quasiperiodic nanoribbons, which were intentionally not included in the fitting.
}

{
The agreement between the fitted lines and the data points even for the largest systems shows the remarkable predictive power of Eq. (\ref{eq:k-vs-L}), which can be used to predict the intrinsic lattice thermal conductivity from simulations with relatively short systems.
According to Fig. \ref{fig03}, simulating nanoribbons of length up to $\approx 100$ nm is enough to predict the intrinsic thermal conductivity of the graphene-hBN superlattices.
The data in Fig. \ref{fig03} also show that for domains of length $d=0.43$ nm, the thermal conductivity decreases with the Fibonacci generation, approaching the conductivity of a random superlattice. 
We attribute this behavior to increased phonon scattering at interfaces, since the number of interfaces per unit cell increases with the generation as illustrated in Fig. \ref{fig01}(b).
This behavior is also related to the decrease in effective phonon mean free path and average group velocity as we shall see in Sec. \ref{sec:33}.
Nonetheless, we notice that even for the eighth Fibonacci generation the thermal conductivity of quasiperiodic superlattices is larger than that of an equivalent random superlattice.
}

\subsection{Dependence of $\kappa$ on domain length}

\begin{figure}[htbp]
\begin{center}
\includegraphics[width=0.8\linewidth]{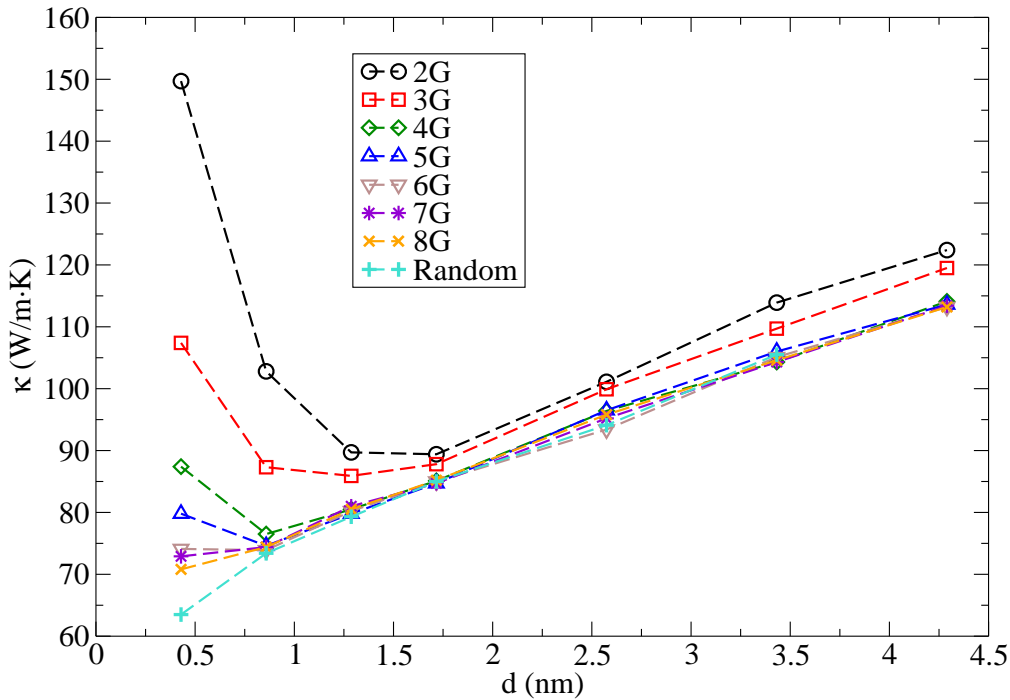}
\caption{Intrinsic thermal conductivity versus domain length. The valley present in some of the curves is associated with the crossover from the coherent to the incoherent heat transport regime, and it disappears from 7G onwards. Dashed lines are guides to the eye.}
\label{fig04}
\end{center}
\end{figure}

{
Now we vary the domain length and analyze its influence on $\kappa$.
Since we want to focus on the effect of the quasiperiodicity on the thermal transport, we consider only superlattices with equal sized domains of graphene and hBN.
Fig. \ref{fig04} shows the intrinsic thermal conductivity as a function of the domain length from the second to the eighth Fibonacci generation. 
The figure also includes the conductivity of random superlattices which were built respecting the two constraints in order to make them equivalent to the quasiperiodic ones in terms of adjacent domains and ratio between domain types.
A non-monotonic behavior of $\kappa$ as a function of $d$, with the presence of a minimum, is a common feature for periodic superlattices such as the second generation of the Fibonacci sequence. 
The presence of a minimum is generally attributed to a crossover from a coherent transport regime to an incoherent one \cite{felix2018}.
}

{
In Fig. \ref{fig04} we notice that the value of $d$ for which $\kappa$ is minimum decreases as  the Fibonacci generation increases.
From 7G onwards, including the random superlattice, $\kappa$ increases monotonically with $d$ and the valley associated with the coherent to incoherent crossover disappears. 
This observation suggests that the presence of quasiperiodicity gradually suppresses coherent thermal transport in graphene-hBN superlattices, approaching a random superlattice in which the heat transport is predominantly incoherent.
The observed suppression is related to the increasing number of interfaces per unit cell, which hinders phonon coherence along the superlattice.
Furthermore, the presence of quasiperiodicity (here from 3G to 8G) decreases the thermal conductivity with respect to the periodic superlattice (2G) with the same domain length $d$.
This is also related to the increasing number of interfaces within the unit cell and to the behavior of the effective phonon mean free path and average group velocity as it will be shown in Sec. \ref{sec:33}.
}

\subsection{Decomposition of the thermal conductivity}

{
Now we try to quantify the coherent thermal transport in Fibonacci quasiperiodic graphene-hBN superlattices.
Following the decomposition of $\kappa$ in coherent and incoherent phonon contributions proposed by Wang et al. \cite{Wang2014}, we can write
\begin{equation}
\kappa = \kappa_{coh} + \kappa_{incoh}.
\end{equation}
In the case of periodic and quasiperiodic superlattices both coherent and incoherent phonons contribute to the thermal conductivity, while in random superlattices only incoherent phonons contribute to $\kappa$ since coherent phonons become localized \cite{Wang2014,Hu2018,Hu2019}.
Therefore, $\kappa_{incoh} = \kappa_{random}$, and we obtain the coherent contribution to the thermal conductivity of a quasiperiodic superlattice by subtracting the conductivity of an equivalent random superlattice:
\begin{equation}
\kappa_{coh} = \kappa - \kappa_{random}.
\end{equation}
}

\begin{figure}[htbp]
\begin{center}
\includegraphics[width=0.8\linewidth]{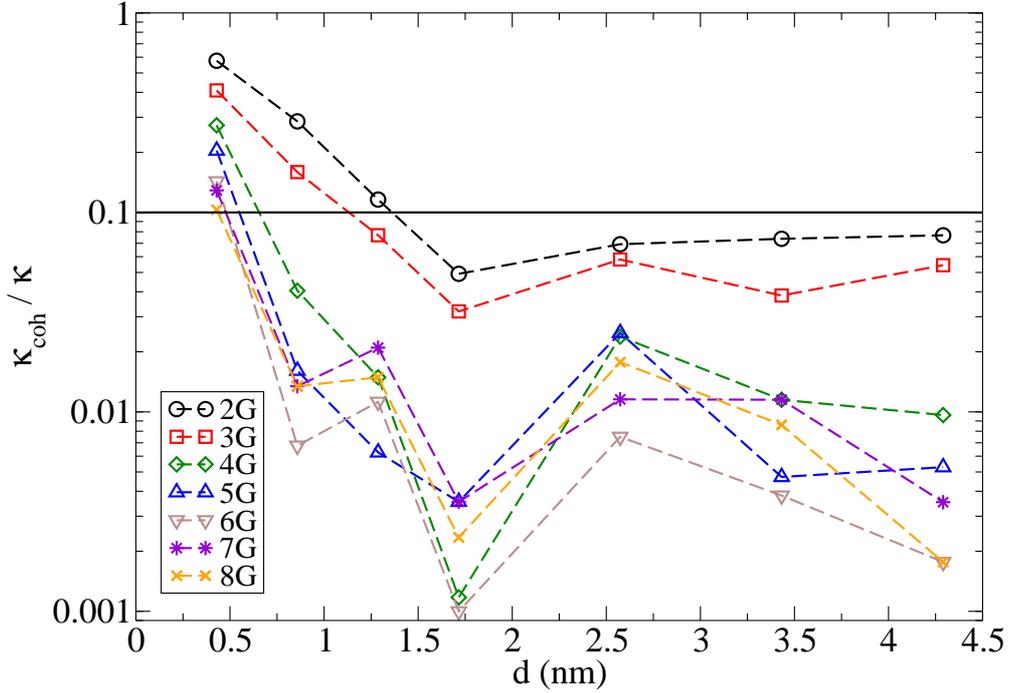}
\caption{Normalized coherent contribution to the thermal conductivity versus domain length. Below the horizontal line incoherent phonons are responsible for more than 90\% of the thermal conductivity.}
\label{fig05}
\end{center}
\end{figure}

{
Fig. \ref{fig05} shows the contribution of coherent transport to the thermal conductivity of quasiperiodic superlattices normalized by their total conductivity.
In superlattices with $d=0.43$ nm the contribution of coherent phonons ranges from $60$\% in the second generation down to $10$\% for the eighth generation.
For $d=0.86$ nm only the second and third generations show a contribution larger than $10$\% from coherent phonons to the total $\kappa$. 
According to Fig. \ref{fig05}, for $d>1.5$ nm at least $90$\% of the thermal conductivity in  quasiperiodic graphene-hBN superlattices it due to incoherent thermal transport.
Therefore, the analysis based on the decomposition of $\kappa$ further corroborates the conclusions drawn from Fig. \ref{fig04} regarding the disappearance of the transition from coherent to incoherent heat transport.
}

We would like to point out that our prediction has measurable implications in future experimental realizations of graphene-hBN superlattices. 
The periodic in-plane heterostructures of graphene and hBN produced by Liu et al. \cite{Liu2013} presented domain lengths ranging from $100$ nm to $1$ $\mu$m.
Due to experimental limitations it is unlikely that superlattices with domain lengths below $10$ nm will be produced in the near future.
{Therefore, according to our simulations the thermal transport in graphene-hBN superlattices experimentally produced should always be in the incoherent regime at room temperature, and their conductivity should increase with the size of the domains.
This feature allows for a direct experimental verification of our predictions at room temperature.}

\subsection{Influence of the Fibonacci generation}
\label{sec:33}

\begin{figure}[htbp]
\begin{center}
\includegraphics[width=0.5\linewidth]{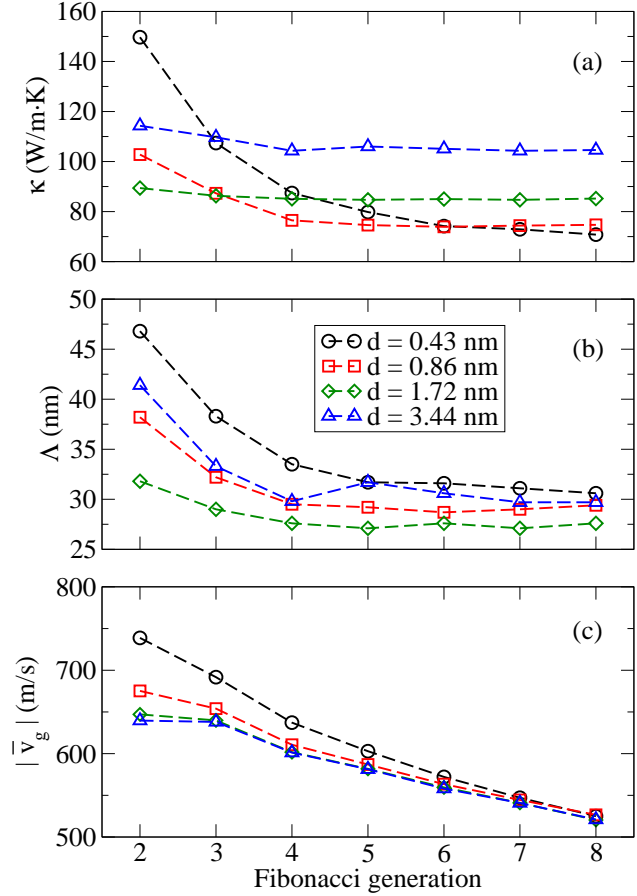}
\caption{(a) Intrinsic thermal conductivity, (b) effective phonon mean free path and (c) average group velocity as functions of Fibonacci generation. The three quantities generally decrease with the generation, but are more sensitive to domain size for earlier generations. Dashed lines are guides to the eye.}
\label{fig06}
\end{center}
\end{figure}

{
Finally, in order to better understand how the quasiperiodicity affects the thermal transport in graphene-hBN superlattice nanoribbons, we consider the influence of the Fibonacci generation on thermal conductivity, phonon mean free path and average group velocity for different domain lengths.
Figures \ref{fig06}(a) and (b) show the intrinsic thermal conductivity and the effective phonon mean free path obtained from Eq. \ref{eq:k-vs-L}.
Meanwhile, in Fig. \ref{fig06}(c) we plot the absolute value of the average phonon group velocity, calculated directly from the phonon dispersion curves.
For the two shortest domain sizes, $0.43$ and $0.86$ nm, the strong decrease of $\Lambda$ and $|\bar{\mathrm{v}}_g|$ as the generation increases explain the decrease in $\kappa$ observed here in panel (a), and also visible in Figs. \ref{fig03} and \ref{fig04}. 
According to the data, the dependence of $\kappa$ on the generation appears to be stronger in the coherent transport regime and weaker in the incoherent one.
}

{
In the case of $d=0.86$ nm, Fig. \ref{fig06} shows a smaller variation of $\kappa$ and $\Lambda$ beyond the third generation, and according to Fig. \ref{fig05} the thermal transport is mostly incoherent after the third generation for that domain length.
A similar behavior is observed for $d=1.72$ nm and $d=3.44$ nm where there is a weak dependence of $\kappa$ on the generation.
This is an indication that, once the system is in the incoherent transport regime, the increasing density of interfaces has a weaker impact on phonon transport.
According to Fig. \ref{fig06}(b), $\Lambda$ decreases with generation in the coherent regime and  oscillates around an average value in the incoherent one. 
Meanwhile the average group velocity decreases with generation for all domain lengths, and becomes independent of $d$ for higher generations.
}

\section{Conclusions}

A final remark must be made regarding the length scales in question.
According to Fig. \ref{fig04}, the domain lengths at which the coherent to incoherent crossover takes place range from $0.43$ to $1.72$ nm.
Thus, the superlattice periods at which the crossover takes place range from $3.44$ nm for the second generation (corresponding to $2$ domains of $1.72$ nm) up to $11.2$ nm for the sixth generation (corresponding to $13$ domains of length $0.86$ nm each).
Now, according to Fig. \ref{fig06}(b), the corresponding effective phonon MFP would be $32$ nm for 2G and approximately $30$ nm for 6G.
Therefore, in the 2G case $\ell_p$ is one order of magnitude smaller than $\Lambda$, while in the 6G case the quantities are actually commensurate.
It has been argued that the crossover occurs when $\ell_p$ approaches the MFP of the major heat carrying phonons although some authors disagree \cite{simkin2000, Latour2014}.
Unfortunately, the effective MFP we obtain from Eq. (\ref{eq:k-vs-L}) corresponds to an average over all phonon modes, whose MFP vary by a few orders of magnitude.
Consequently, our analysis does not enable a definite conclusion regarding which length scales are indeed responsible for the crossover.

In summary, we performed large scale non-equilibrium MD simulations to systematically investigate the thermal transport properties of quasiperiodic graphene-hBN nanoribbons. 
We observed that the length of individual domains for which the thermal conductivity reaches a minimum decreases as the Fibonacci generation increases.
Since the presence of a valley in the thermal conductivity is associated with a transition from a coherent transport regime to an incoherent one, we conclude that the presence of a quasiperiodicity hinders coherent phonon transport in these superlattices.
Based on simulation results, we conclude that from the seventh Fibonacci generation and beyond the heat transport in graphene-hBN superlattices is mostly incoherent for any realizable domain length at room temperature.
This observation could, at least in principle, be measured in experiments at room temperature with current fabrication techniques.
The suppression of coherent thermal transport in graphene-hBN superlattices enables a further degree of control on heat conduction at the nanoscale, which has the potential to improve the design of novel thermal management devices.

\section*{Acknowledgements}
The authors thank Davide Donadio, Mauro S. Ferreira and Zheyong Fan for a critical reading of the manuscript.
We acknowledge financial support from Coordena\c{c}\~ao de Aperfei\c{c}oamento de Pessoal de N\'{\i}vel Superior (CAPES), and Conselho Nacional de Desenvolvimento Cient\'{\i}fico e Tecnol\'ogico (CNPq) (Grants 309961/2017 and  436859/2018).
Computational resources were provided by the High Performance Computing Center (NPAD) and the CLIMA cluster at UFRN.

\section*{References}


\providecommand{\noopsort}[1]{}\providecommand{\singleletter}[1]{#1}%

\end{document}